\documentclass[aps,prc,twocolumn,showpacs,preprintnumbers,nofootinbib,float,superscriptaddress,longbibliography]{revtex4-2}

\usepackage{graphicx, fancybox}
\usepackage{amsmath,amssymb}
\usepackage[colorlinks=true, pdfstartview=FitV, linkcolor=red, citecolor=blue, urlcolor=blue]{hyperref}
\usepackage{color}
\usepackage{xcolor}
\usepackage{soul}
\usepackage{array}
\usepackage{url}
\usepackage[utf8]{inputenc}
\usepackage{multirow}
\usepackage{graphicx}

\newcommand{\snn}{\sqrt{s_\mathrm{NN}}}
\newcommand{\pT}{p_\perp}

\usepackage[normalem]{ulem}

\begin{document}

\title{High-order anisotropic flow and nonlinear hydrodynamic responses}

\author{Owen Horecny}
\email{hi6127@wayne.edu}
\affiliation{Department of Physics and Astronomy, Wayne State University, Detroit, Michigan, 48201, USA}

\author{Chun Shen}
\email{chunshen@wayne.edu}
\affiliation{Department of Physics and Astronomy, Wayne State University, Detroit, Michigan, 48201, USA}

\begin{abstract}
In this work, we study the charged hadron anisotropic flow coefficients $v_n$ (up to $n = 12$) using high-statistics event-by-event simulations of Pb+Pb collisions at $\snn = 5.02$ and $5.36$~TeV, employing the IP-Glasma + MUSIC + UrQMD hybrid approach. The power spectra of anisotropic flow coefficients ($v_n\{2\}$ vs. $n$) are compared with the ALICE measurements from central to 50\% centrality. To understand the various sources contributing to the high-order $v_n$ coefficients, we analyze the nonlinear mode coefficients for high-order anisotropic flow coefficients using different approximations and make comparisons with available measurements. 
\end{abstract}

{\maketitle}

\section{Introduction}

High-energy relativistic nuclear experiments at the Relativistic Heavy-Ion Collider (RHIC) and the Large Hadron Collider (LHC) have given us insight into the formation of nuclear matter in extreme conditions, i.e., the Quark Gluon Plasma (QGP)~\cite{Shuryak:2014zxa, Achenbach:2023pba, Arslandok:2023utm, Shen:2020gef, Shen:2020mgh, Gale:2013da, Heinz:2013th}. In general, the dynamical evolution of a heavy-ion collision goes through three distinct phases: an early-stage pre-equilibrium phase where the system expands primarily in the longitudinal direction and preserves its transverse geometry while evolving towards chemical and thermal equilibrium~\cite{Schlichting:2019abc, Berges:2020fwq}; a hydrodynamic phase where local pressure gradients drive the medium's expansion and convert the fireball's shape to flow velocity distribution~\cite{Ollitrault:1992bk, Heinz:2004qz, Romatschke:2009im}; and a ``freeze-out'' phase where the QGP undergoes hadronization and falling apart from each other~\cite{Hirano:2005xf, Petersen:2008dd, Shen:2014vra}. Experimental measurements in high-energy heavy-ion collisions have demonstrated that the strongly coupled nature of QGP converts the local pressure gradients into anisotropic flows, driven by the geometric shape of the interaction zone. These flow patterns imprint themselves on the azimuthal variations of the momentum distributions of final-state hadrons. In analogy to how primordial quantum fluctuations evolved in our early universe via inflation, hydrodynamic expansion translates the initial shape fluctuations of heavy-ion collisions into a power spectrum of anisotropic flow coefficients for final-state hadrons.

The event-by-event anisotropic flow coefficients $\{V_n\}$ are defined via a Fourier expansion of the azimuthal distribution of the final-state particles,
\begin{align}
     \frac{dN}{dy \pT d\pT d\phi}=\frac{1}{2\pi} \frac{dN}{dy \pT d\pT} \sum_{n=-\infty}^{n=+\infty} V_n e^{-i n \phi},
     \label{eq:FourierVn}
\end{align}
where the $n$-th order anisotropic flow coefficient $V_n$ is a complex number, $V_n \equiv v_n \exp(i n \Psi_n)$ with the real $v_n$ and $\Psi_n$ as its magnitude and phase, respectively. From the definition, we have $V_{-n} = V_n^*$. Because the impact parameters of collision events can not be measured and are randomly oriented in the transverse plane event by event, the event-averaged anisotropic flow vector $\langle V_n \rangle = 0$ for an ensemble of collision events. Experiments measure $v_n$ via multi-particle correlations in which the global phase among the flow vectors cancel each other, e.g., $C_n\{2\} = \langle V_n V_n^* \rangle$ and $C_{mnk}\{3\} = \langle V_m V_n V_k^* \rangle$ with $m + n = k$~\cite{Qiu:2012uy, Luzum:2012da}.

The elliptic and triangular flow coefficients $V_2$ and $V_3$ show strong correlations with the second and third-order deformation of the initial density profile characterized by initial eccentricity $\mathcal{E}_n$ defined as~\cite{Teaney:2010vd, Qiu:2011iv, Gardim:2011xv}
\begin{align}
    \mathcal{E}_n = - \frac{\int d^2 r e(r, \phi) e^{i n \phi}}{\int d^2 r e(r, \phi)},
\end{align}
where the weight $e(r, \phi)$ is the initial energy density profile. For harmonic order $n = 2, 3$, the anisotropic flow coefficients show strong linear responses to the collision geometry, $V_n = k_n \mathcal{E}_n$. The high-order anisotropic flow coefficient $V_n$ ($n > 3$) not only exhibits dependence on the $n$-th order eccentricity $\mathcal{E}_n$ of the initial density profile, probing fine details about the medium at small length scales, it also receives contributions from the low-order deformation via nonlinear interactions during the hydrodynamic evolution, serving as a potential probe for the QGP viscosity and speed of sound in the equation of state for dense nuclear matter~\cite{Staig:2011wj, Teaney:2012ke}.

The high statistics runs from the LHC experiments have recently pushed flow correlation measurements to multi-particles, such as $v_2\{10\}$~\cite{CMS-PAS-HIN-21-010, Milosevic:2023dhx}, and high-order flow harmonics. The ALICE Collaboration has recently measured two-particle charged hadron anisotropic flow coefficients $v_n\{2\}$ up to $n = 9$ in Pb-Pb collisions at $\snn = 5.02$ TeV~\cite{ALICE:2020sup}. In this paper, we will perform high-statistics numerical simulations to make theoretical comparisons with these measurements. We will quantify the relative contributions of linear and nonlinear modes to the high-order $v_n$ for $n = 4-9$, providing insight into these measurements.

The rest of the paper is organized as follows. Section~\ref{Sec:analysis} will outline the formalism for analyzing linear and nonlinear contributions in high-order anisotropic flow harmonics. We will then compare our numerical results and ALICE measurements in Sec.~\ref{Sec:results}. We will conclude in Sec.~\ref{Sec:conclusion}.

\section{Linear and nonlinear anisotropic flow responses}
\label{Sec:analysis}

We can generally decompose the complex high-order anisotropic flow coefficients $V_n$ ($n \geq 4$) into linear and nonlinear terms, where the linear term represents a response to the initial geometry deformation at the same order ($\mathcal{E}_n$). The nonlinear modes arise from fluctuations in lower-order harmonics ($m < n$)~\cite{Yan:2015jma}. Let us start with the decomposition of $V_4$ and $V_5$ as follows,
\begin{align}
    V_4 &= V_{4L} + \chi_{4,22} V_2^2 \label{eq:v4} \\
    V_5 &= V_{5L} + \chi_{5,23} V_2 V_3, \label{eq:v5}
\end{align}
where $V_{4L}$ and $V_{5L}$ are the linear contributions, $\chi_{4,22}$ and $\chi_{5,23}$ are the nonlinear mode coefficients. The linear flow coefficients are defined to be orthogonal to the nonlinear terms,
\begin{align}
    \langle V_{4L} (V_2^2)^* \rangle \equiv 0 \mbox{ and } \langle V_{5L} (V_2 V_3)^* \rangle \equiv 0.
\end{align}
Using these conditions, we can obtain the nonlinear mode coefficients as follows,
\begin{align}
    \chi_{4,22} &= \frac{\left\langle V_4(V_2^*)^2\right\rangle}{\left\langle |V_2|^4 \right\rangle} \label{eq:chi422} \\
    \chi_{5,23} &= \frac{\left\langle V_5V_2^*V_3^*\right\rangle}{\left\langle |V_2|^2|V_3|^2\right\rangle}. \label{eq:chi523}
\end{align}
Plugging the values of Eqs.~\eqref{eq:chi422} and \eqref{eq:chi523} into Eqs.~\eqref{eq:v4} and \eqref{eq:v5}, respectively, we obtain the event-by-event linear flow vectors,
\begin{align}
    V_{4L} &= V_4 - \chi_{4,22} V_2^2  \label{eq:V4L} \\
    V_{5L} &= V_5 - \chi_{5,23} V_2 V_3. \label{eq:V5L}
\end{align}
Finally, we can compute the RMS of the linear flow coefficients $\sqrt{\langle |V_{4L}|^2 \rangle}$ and $\sqrt{\langle |V_{5L}|^2 \rangle}$ over an ensemble of collision events.

For higher-order anisotropic flow coefficients $V_n$ with $n\geq6$, their decompositions are more complex, and they read as follows~\cite{Qian:2016fpi, McDonald:2017ayb, Giacalone:2018wpp},
\begin{align}
    V_6 &= V_{6L} + \chi_{6,222}V_2^3 + \chi_{6,33}V_3^2 + \chi_{6,24}V_2V_{4L} \label{eq:v6} \\
    V_7 &= V_{7L} + \chi_{7,223}V_2^2V_3 \nonumber \\
        & \quad + \chi_{7,25}V_2V_{5L} + \chi_{7,34}V_3V_{4L} \label{eq:v7} \\
    V_8 &= V_{8L}+\chi_{8,2222}V_2^4 +\chi_{8,233}V_2V_3^2 +\chi_{8,224}V_2^2V_{4L}  \nonumber \\
        & \quad +\chi_{8,26}V_2V_{6L}+\chi_{8,35}V_3V_{5L}+\chi_{8,44}V_{4L}^2 \label{eq:v8} \\
    V_9 &= V_{9L} + \chi_{9,2223}V_2^3V_3 + \chi_{9,333}V_3^3  \nonumber \\
        &\quad + \chi_{9,234}V_2V_3V_{4L} + \chi_{9,225}V_2^2V_{5L} \nonumber \\
        & \quad + \chi_{9,27}V_2V_{7L} + \chi_{9,36}V_3V_{6L} + \chi_{9,45}V_{4L}V_{5L}. \label{eq:v9}
\end{align}
Again, we define the linear flow coefficients as independent of all the nonlinear terms in the decomposition. Take $V_6$ as an example,
\begin{align}
    \langle V_{6L} (V_2^3)^* \rangle \equiv 0, \langle V_{6L} (V_3^2)^* \rangle \equiv 0, \langle V_{6L} (V_2 V_{4L})^* \rangle \equiv 0.
\end{align}
With these conditions, we can solve the nonlinear mode coefficients as a matrix equation as follows,
\begin{widetext}
\begin{align}
\renewcommand\arraystretch{1.8}
    \begin{pmatrix}
            \left\langle (V_2^3)^* V_6 \right\rangle\\
            \left\langle (V_3^2)^* V_6 \right\rangle\\
            \left\langle V_2^*V_{4L}^*V_6\right\rangle \\
    \end{pmatrix}
    =
    \begin{pmatrix}
            \left\langle|V_2|^6 \right\rangle & \left\langle (V_2^3)^* V_3^2 \right\rangle & \left\langle (V_2^3)^* V_2 V_{4L} \right\rangle\\
            \left\langle (V_3^2)^* V_2^3 \right\rangle & \left\langle |V_3|^4 \right\rangle & \left\langle (V_3^2)^* V_2V_{4L} \right\rangle\\
            \left\langle V_2^*V_{4L}^*V_2^3 \right\rangle & \left\langle V_2^*V_{4L}^*V_3^2 \right\rangle & \left\langle |V_2|^2|V_{4L}|^2 \right\rangle
    \end{pmatrix}
    \begin{pmatrix}
            \chi_{6,222} \\
            \chi_{6,33} \\
            \chi_{6,24}
    \end{pmatrix}. \label{eq:chi6_MatrixEq}
\end{align}      
\end{widetext}
Once we solve all three nonlinear mode coefficients together, we can compute the event-by-event linear flow vector as
\begin{align}
    V_{6L} = V_6 - \chi_{6,222} V_2^3 - \chi_{6,33} V_3^2 - \chi_{6,24} V_2 V_{4L}.
    \label{eq:V6L}
\end{align}
Similarly, based on Eqs.~\eqref{eq:v7}, \eqref{eq:v8}, and \eqref{eq:v9}, we can construct matrix equations for $V_7$ to $V_9$.

All the flow correlations in Eq.~\eqref{eq:chi6_MatrixEq} are real after event averaging because one can swap the particle pairs in the flow vectors and their complex conjugate. We expect the nonlinear mode coefficients $\chi$ to be real in general. We verified numerically that all the imaginary parts of the nonlinear mode coefficients are consistent with zero within statistical fluctuations.

One may wonder why the nonlinear terms do not include the directed flow $V_1$ in the decompositions. In Appendix~\ref{sec:V1}, we will show that the directed flow coefficient decouples from all the higher-order anisotropic flow coefficients. Similar results were demonstrated in Ref.~\cite{Qian:2016fpi}. Its related nonlinear terms have negligible contributions in our analysis.

We note that the nonlinear mode coefficients measured by the ALICE Collaboration were defined as~\cite {ALICE:2020sup},
\begin{align}
    &\chi'_{6,222} = \frac{\langle V_6 (V_2^3)^* \rangle}{\langle |V_2|^6 \rangle},& &\chi'_{6,33} = \frac{\langle V_6 (V_3^2)^* \rangle}{\langle |V_3|^4 \rangle}, \label{eq:ALICEchi6} \\
    &\chi'_{6,24} = \frac{\langle V_6 (V_2 V_{4L})^* \rangle}{\langle |V_2|^2 |V_{4L}|^2 \rangle},&
    &\chi'_{7,223} = \frac{\langle V_7 (V_2^2 V_3)^* \rangle}{\langle |V_2|^4 |V_3|^2 \rangle}. \label{eq:ALICEchi7}
\end{align}
They correspond to the solutions when ignoring the off-diagonal elements in the matrix equations for $V_6$ (in Eq.~\eqref{eq:chi6_MatrixEq}) and $V_7$. In the next section, we will quantify the effects of including off-diagonal elements on the extracted nonlinear mode coefficients.

\section{Phenomenological study at the LHC}
\label{Sec:results}

In this work, we employ the IP-Glasma + MUSIC + UrQMD hybrid framework~\cite{Paquet:2015lta, Schenke:2020mbo, Schenke:2020unx, Mantysaari:2024uwn} to study the high-order anisotropic flow coefficients and their decompositions into the linear flow and nonlinear mode coefficients. All the model parameters are specified in Ref.~\cite{Mantysaari:2024uwn}.
We simulate 300k hydrodynamic events for minimum bias Pb+Pb collisions at $\snn = 5.02$~TeV. 

Because the UrQMD transport model~\cite{Bass:1998ca, Bleicher:1999xi} simulates the hadronic dynamics with the Monte-Carlo method, we construct the anisotropic flow vectors of individual hydrodynamic events with the final-state charged hadrons from multiple (oversampled) hadronic events~\cite{McDonald:2016vlt},
\begin{align}
    V_n =\frac{1}{N}\sum_{j=1}^N e^{in\phi_j},
\end{align}
where the summation runs over all final-state charged hadrons from the UrQMD events sampled from the same hydrodynamic hypersurface~\cite{Shen:2014vra}. The average number of oversampled hadronic events is about 100 per hydrodynamic event. Because our simulations assume longitudinal boost invariance, we include particles within a large rapidity region to compute the anisotropic flow vectors, thereby increasing statistics further. We checked that the $v_n\{2\}$ results are within a 2\% difference between flow vectors from $| \eta | \in [0.4, 0.8]$ and those with $| \eta | \in [0.4, 3.2]$. At the same time, the latter rapidity cut increases the statistics by a factor of 49 for two-particle correlations.

In every hydrodynamic event, we compute two sub-event anisotropic flow vectors, namely $\{V_n^A\}$ with kinematic cuts $p_T \in [0.2, 3]$~GeV and $\eta \in [-3.2, -0.4]$ and $\{V_n^B\}$ using charged hadrons with $p_T \in [0.2, 3]$~GeV and $\eta \in [0.4, 3.2]$.
We note that the ALICE experiment measured $v_n\{2\}$ with $p_T \in [0.2, 5]$ GeV. We have verified that the relative difference of $v_n\{2\}$ is within 4\% of our model's prediction between the two $p_T$ cuts, within our statistical uncertainties.
We compute $v_n\{2\}$ as
\begin{align}
    v_n\{2\} &= \sqrt{\Re\{\langle V_n^A (V_n^B)^* \rangle\}} \nonumber \\
    &= \sqrt{\frac{1}{2}\langle V_n^A (V_n^B)^* + (V_n^A)^* V_n^B \rangle}.
\end{align}
The inner product of the two complex vectors imposes a rapidity gap of $| \Delta \eta | \ge 0.8$ for all particle pairs. The equation on the second line doubles the statistics.
When computing multiple-particle correlations (with more than two particles) for the nonlinear mode coefficients, we use one subevent flow vector for all the complex conjugate terms and the other for the normal flow vectors. The oversampling factor suppresses self-correlations, which have negligible effects on the multi-particle correlations.

Appendix~\ref{sec:numericalConvergence} presents a few numerical tests which verify that the simulated high-order $v_n\{2\}$ coefficients have less than 10\% relative numerical uncertainty.

\subsection{The anisotropic flow power spectrum}

The power spectrum of temperature fluctuations is a powerful tool for studying cosmic microwave background (CMB) radiation~\cite{Planck:2019nip}. 
In central heavy-ion collisions, all the anisotropic harmonic flow coefficients are driven by event-by-event fluctuations, where all orders of initial eccentricity $|\mathcal{E}_n|$ are comparable in their sizes~\cite{Heinz:2013th}.
Refs.~\cite{Staig:2010pn, Staig:2011wj, Lacey:2013is, Shuryak:2017aol} proposed that the eccentricity scaled anisotropic flow $v_n/|\mathcal{E}_n| \propto \exp[- bn^2]$ with the coefficient $b$ depending on the specific viscosity used in the hydrodynamic simulations. In contrast, Ref.~\cite{Hatta:2015era, Hatta:2016czn} showed that the $v_n/|\mathcal{E}_n|$ decays exponentially at large $n$.

\begin{figure}[h!]
    \centering
    \includegraphics[width=\linewidth]{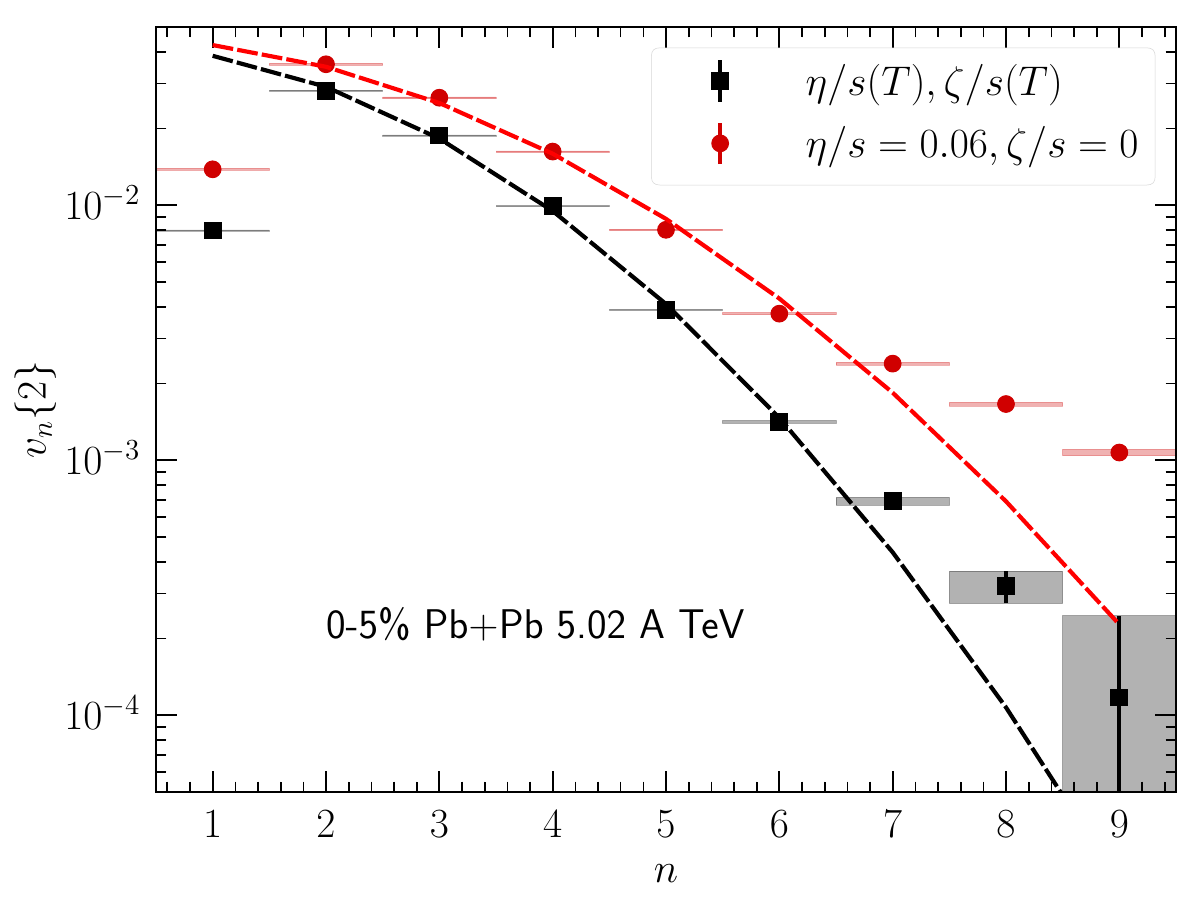}
    \caption{The power spectrum of anisotropic flow coefficients $v_n\{2\}$ with two choices of specific viscosity used in the hydrodynamic simulations. The dashed lines are fits with the functional form, $v_n\{2\}(n) = a e^{- b n^2}$.}
    \label{fig:vnPowerSpectrum}
\end{figure}

Figure~\ref{fig:vnPowerSpectrum} shows the $v_n\{2\}$ power spectrum in 0-5\% Pb+Pb collisions at $\snn = 5.02$~TeV. In addition to the default simulations with temperature-dependent shear and bulk viscosity calibrated in Ref.~\cite{Mantysaari:2024uwn}, we also show a reference result using a constant $\eta/s = 0.06$ and $\zeta/s = 0$ for comparison. The power spectrum in Fig.~\ref{fig:vnPowerSpectrum} shows that the high-order $v_n\{2\}$ coefficients have strong sensitivity to the viscosity used in the simulations.

In central Pb+Pb collisions, we can assume the magnitudes of initial eccentricity $|\mathcal{E}_n|$ are roughly the same for all order $n$~\cite{Heinz:2013th}. The anisotropic flow coefficients $v_2\{2\}$ to $v_6\{2\}$ from our model show the $\exp[- b n^2]$ dependence for both cases, with the coefficient $b$ is smaller for the $\eta/s = 0.06, \zeta/s = 0$ case, qualitatively agreeing with the proposed relation from Refs.~\cite{Staig:2010pn, Staig:2011wj, Lacey:2013is, Shuryak:2017aol}. In the meantime, the high-order $v_n\{2\}$ with $n \ge 7$ decays exponentially with $n$, consistent with the expectations from Refs.~\cite{Hatta:2015era, Hatta:2016czn}.

In the following studies, we present results using temperature-dependent shear and bulk viscosity, calibrated in Ref.~\cite{Mantysaari:2024uwn}.

\begin{figure}[t!]
    \centering
    \includegraphics[width=\linewidth]{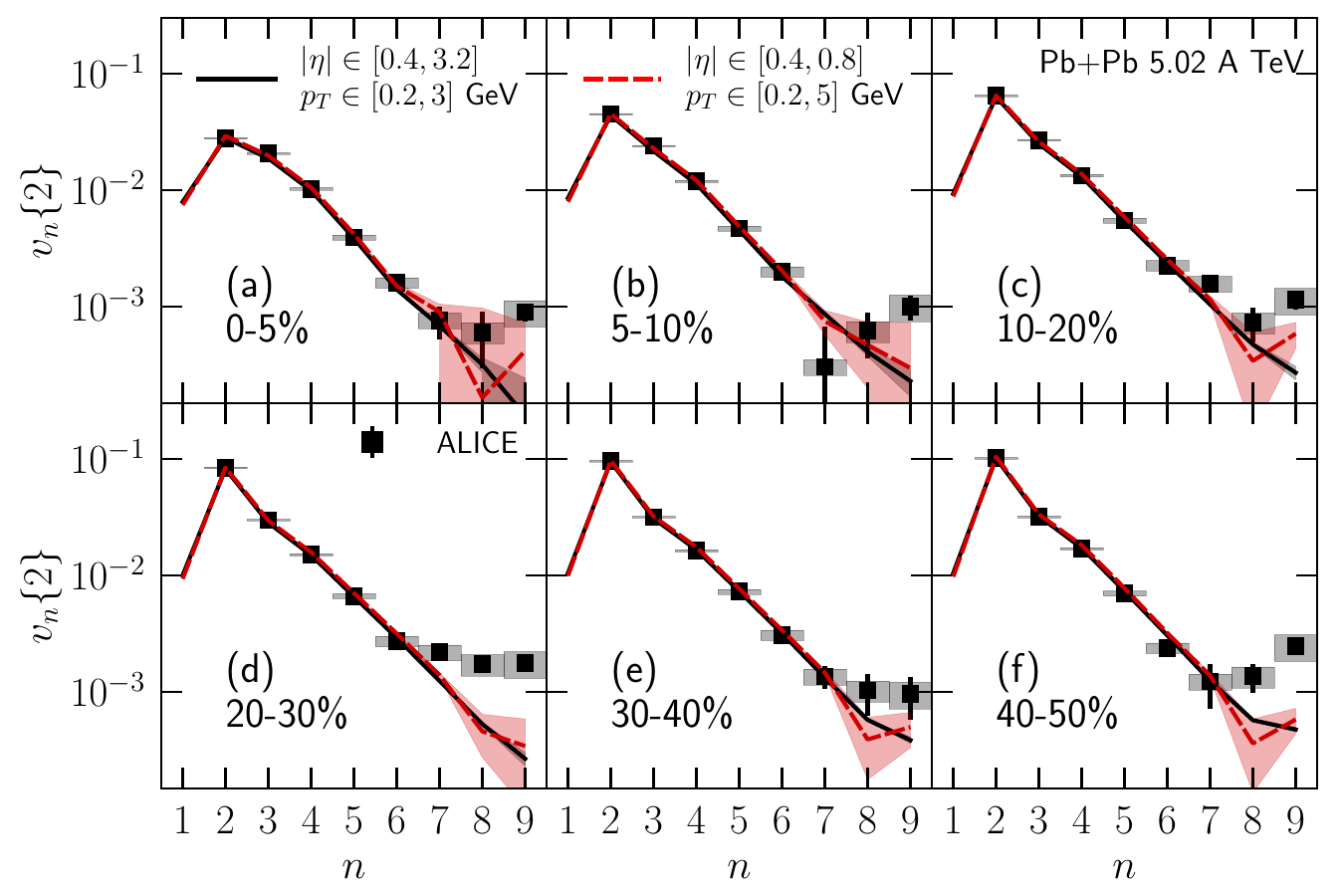}
    \caption{Charged hadron anisotropic flow coefficients $v_n\{2\} (n = 2-9)$ in Pb+Pb collisions at $\snn = 5.02$~TeV with two kinematic cuts from the boost-invariant simulations compared with the ALICE measurements~\cite{ALICE:2020sup}. Individual panels present the results in different centrality bins from central up to 50\%.}
    \label{fig:vn2_vs_n}
\end{figure}

\begin{figure}[t!]
    \centering
    \includegraphics[width=\linewidth]{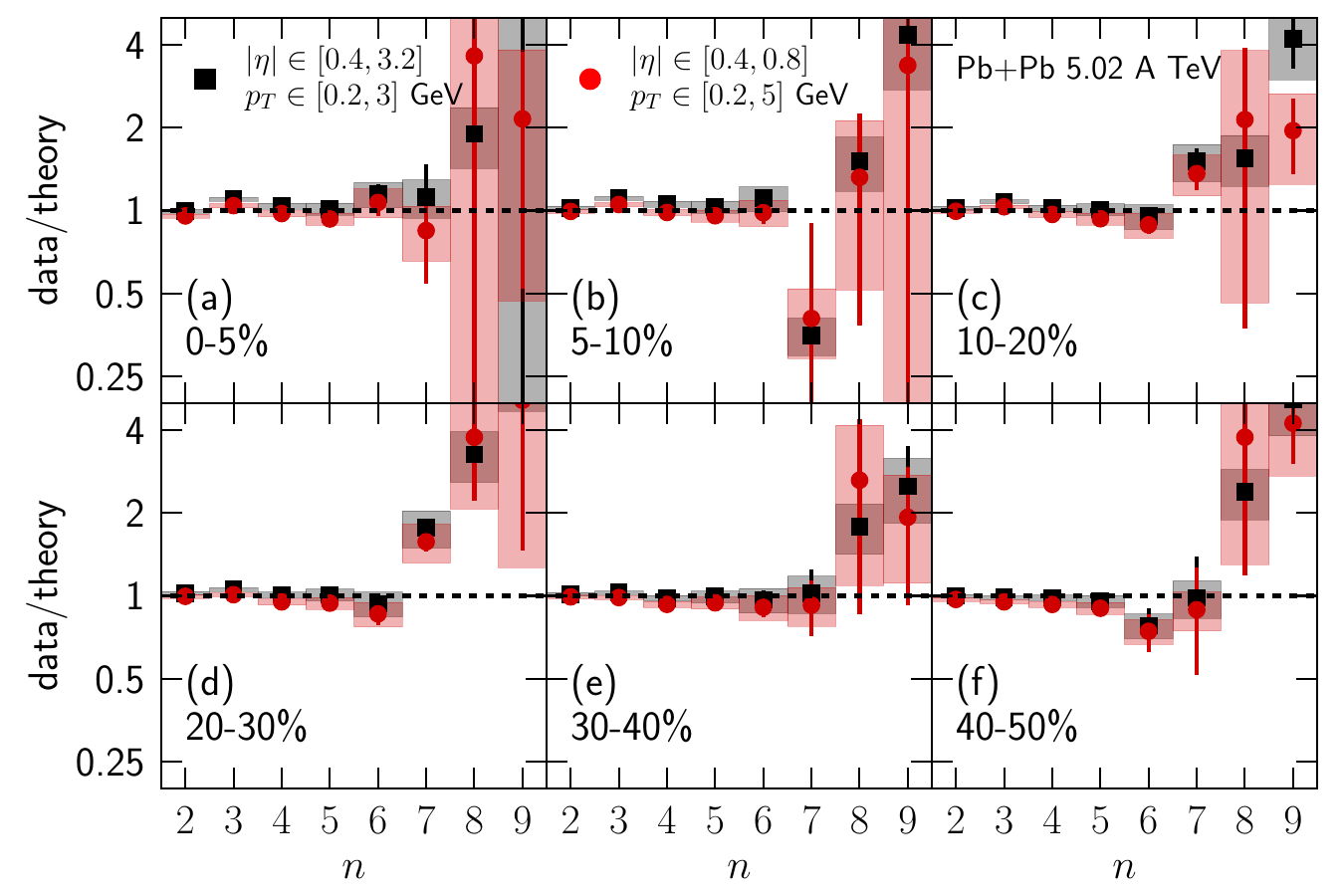}
    \caption{The charged hadron $v_n\{2\}$ ratios for the ALICE data over the theory simulations as functions of the harmonic order from Fig.~\ref{fig:vn2_vs_n}.}
    \label{fig:vn2Ratio}
\end{figure}

\begin{figure}[t!]
    \centering
    \includegraphics[width=\linewidth]{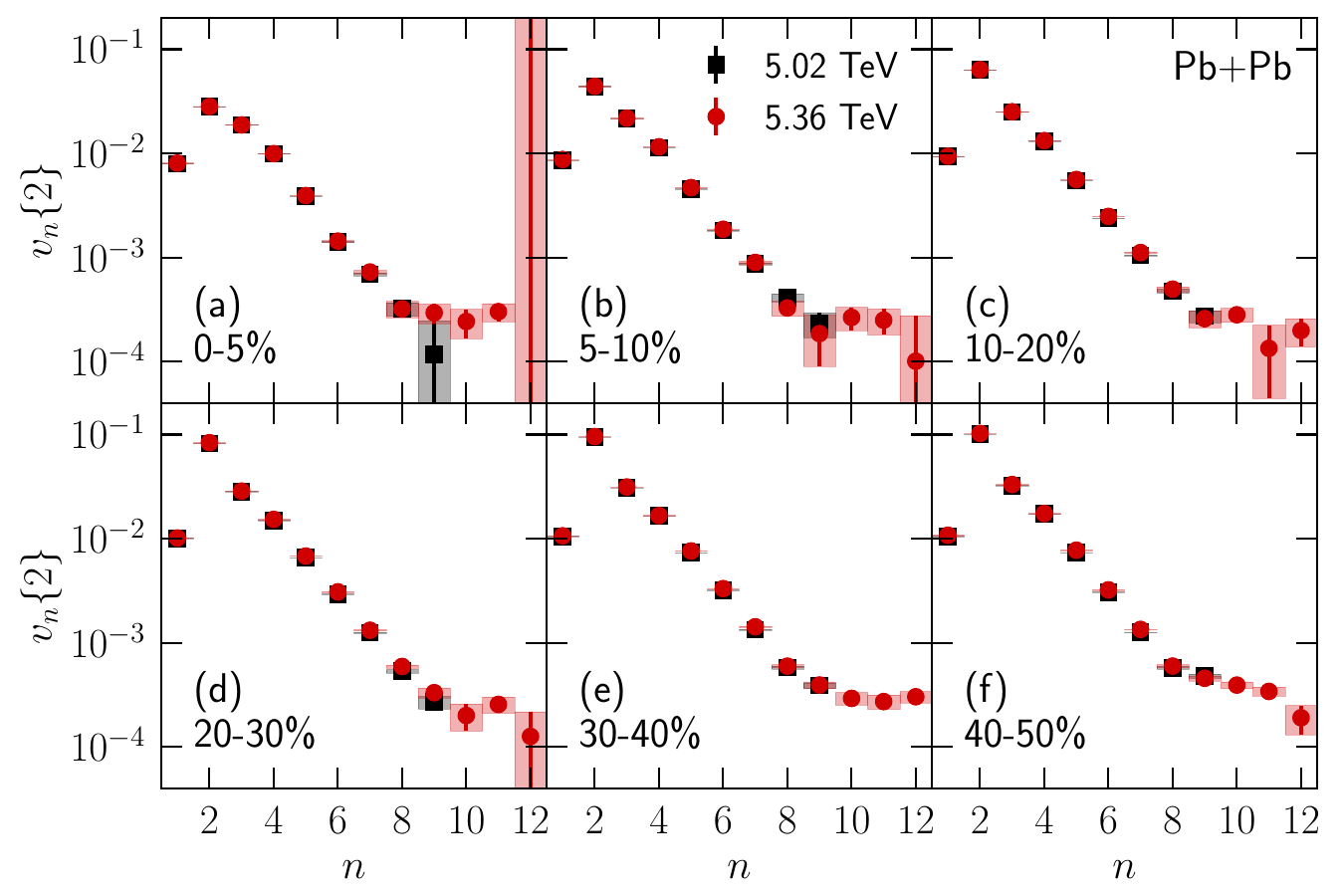}
    \caption{Model prediction for charged hadron $v_n\{2\}$ as a function of $n$ $(n = 1 - 12)$ in Pb+Pb collisions at $\snn = 5.36$~TeV.}
    \label{fig:vn2pred}
\end{figure}

Figure~\ref{fig:vn2_vs_n} shows the comparison with the ALICE measurement on anisotropic flow coefficients $v_n\{2\}$ as functions of the harmonic order $n$ for six centrality bins in Pb+Pb collisions at $\snn = 5.02$~TeV.
Our boost-invariant simulations yield compatible $v_n\{2\}$ results with the different kinematic cuts. The results with the large rapidity acceptance have smaller statistical uncertainties.
Figure~\ref{fig:vn2Ratio} shows the ratios between the experimental measurements and the theoretical calculations. We note that the model calibration performed in Ref.~\cite{Mantysaari:2024uwn} only focused on $v_2\{2\}$ measurements. Remarkably, the IP-Glasma initial conditions can provide a simultaneous and accurate description of the ALICE measurements from $v_2\{2\}$ to $v_7\{2\}$ across all centrality bins. This comparison demonstrates that the IP-Glasma model captures initial-state fluctuations over a wide range of scales~\cite{Schenke:2018fci}. 
Looking at higher-order flow coefficients, we observe that the ALICE measured $v_8\{2\}$ and $v_9\{2\}$ saturate around $10^{-3}$, while our theoretical results continue to decay exponentially. 

Figure~\ref{fig:vn2pred} shows our model predictions for the anisotropic flow power spectra in Pb+Pb at $\snn = 5.36$ TeV. Based on the numerical convergent tests shown in Appendix~\ref{sec:numericalConvergence}, we estimate that the extrapolated numerical uncertainties for the higher-order coefficients  $v_{10}\{2\}$ to $v_{12}\{2\}$ should be comparable with their statistical errors shown in Fig.~\ref{fig:vn2pred}. Therefore, we extend our model prediction for the anisotropic flow power spectrum up to $v_{12}\{2\}$. The results for $v_2\{2\}$ to $v_9\{2\}$ are within 5\% difference to those at 5.02 TeV. We observe that the high-order $v_n\{2\}$ coefficients with $n \ge 10$ saturate their values around $10^{-4}$ in our simulations.

\subsection{Anisotropic flow mode decomposition for $v_4-v_9$}

In this section, we perform mode decomposition for anisotropic flow coefficients $V_4$ to $V_9$ in Pb+Pb collisions at $\snn = 5.02$~TeV.

\begin{figure}[h!]
    \centering
    \includegraphics[width=\linewidth]{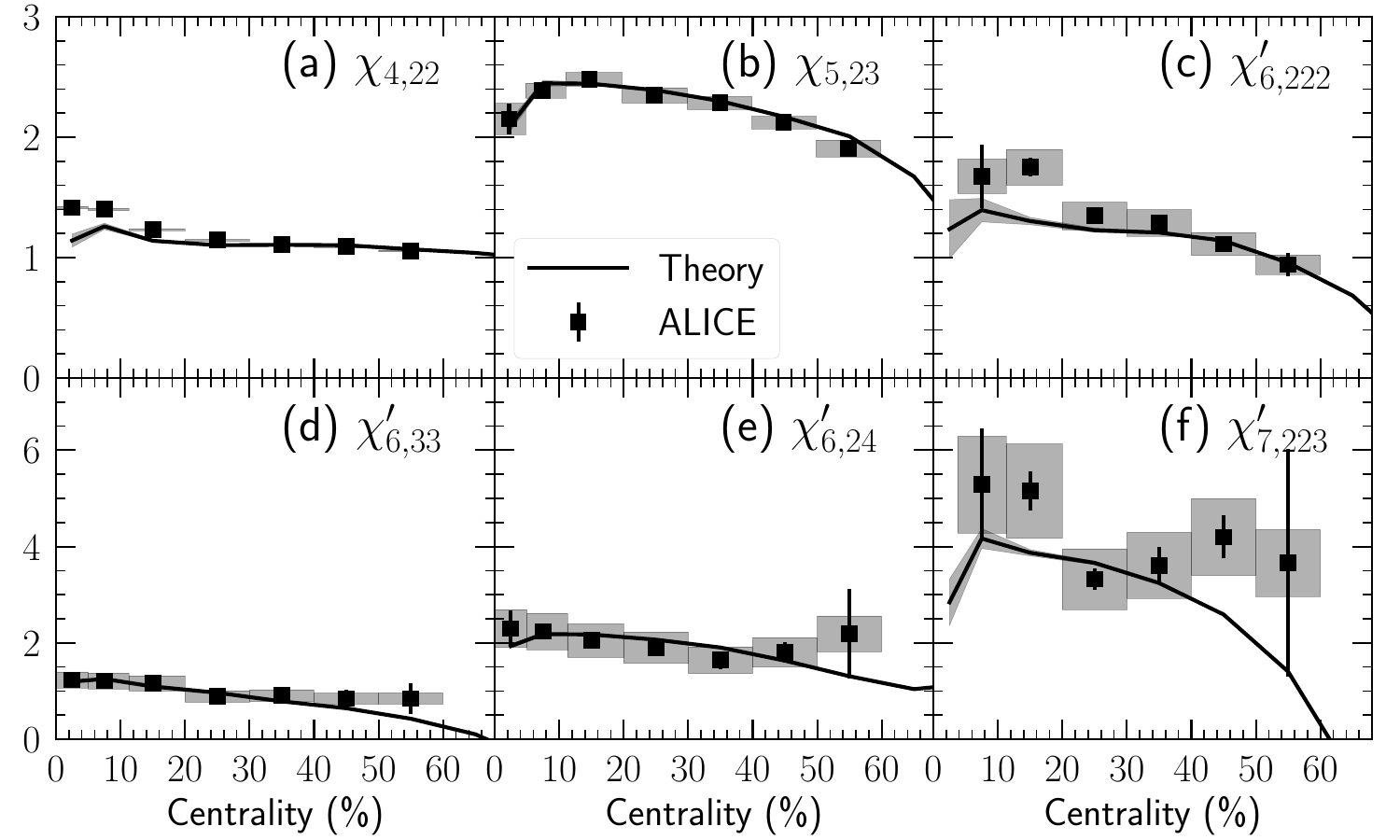}
    \caption{Nonlinear mode coefficients for high-order anisotropic flow coefficients compared with the ALICE measurements in Pb+Pb collisions at $\snn = 5.02$~TeV~\cite{ALICE:2020sup}.}
    \label{fig:chivsALICE}
\end{figure}

Figure~\ref{fig:chivsALICE} shows the nonlinear mode coefficients compared with the ALICE measurements~\cite{ALICE:2020sup} for several centrality bins in Pb+Pb collisions at $\snn = 5.02$~TeV. For apples to apples comparisons, we compute the nonlinear mode coefficients according to Eqs.~\eqref{eq:ALICEchi6} and \eqref{eq:ALICEchi7} for $V_6$ and $V_7$. Overall, we find a good agreement between our model results and the ALICE measurements, demonstrating our hybrid model as a precision tool for these observables.

We note that our model underestimates the values of $\chi_{4,22}$ by 15\% in 0-10\% central Pb+Pb collisions, where the ALICE data rises faster than our model. Based on a detailed comparison with the ALICE measurements on $v_2\{2\}$, $v_4\{2\}$, and $v_{4,22}$ observables, we find that the difference in $\chi_{4,22}$ mainly comes from the correlation $\langle V_4 (V_2^2)^* \rangle$ in central collisions, suggesting that the comparison can be improved if we allow for a non-zero elliptical ($\beta_2$) deformation of the Pb nucleus.

Compared to the previous IP-Glasma + MUSIC results in Refs.~\cite{McDonald:2016vlt, ALICE:2020sup}, our current model provides a better overall description of the non-linear mode coefficients because the current initial-state model includes sub-nucleonic structures constrained by the HERA diffractive $J/\Psi$ measurements, and it leads to different optimal QGP-specific shear and bulk viscosities~\cite{Mantysaari:2024uwn}.

As we have pointed out in the previous section, the nonlinear mode coefficients ALICE measured for $V_6$ and $V_7$ neglected the off-diagonal matrix elements in the decomposition procedure. 

\begin{figure}[h!]
    \centering
    \includegraphics[width=\linewidth]{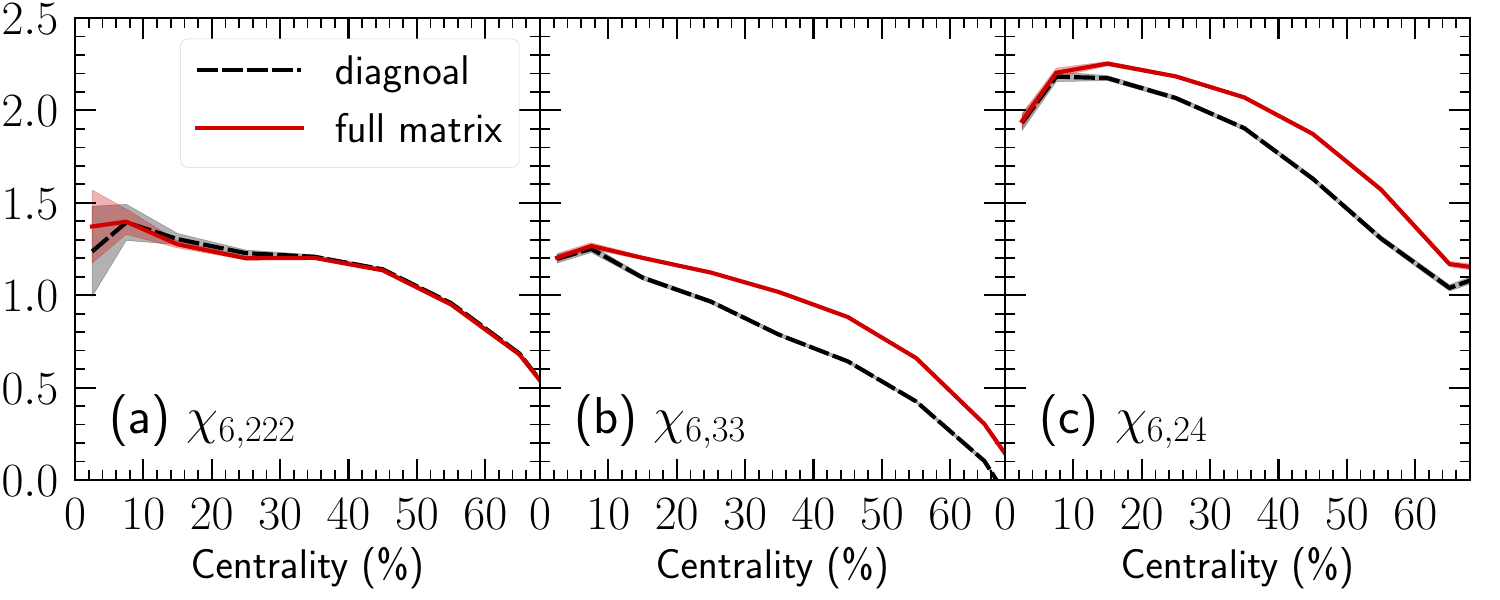}
    \caption{The nonlinear mode coefficients for $v_6$ by solving the full matrix equation vs. only including the diagonal elements.}
    \label{fig:chi6}
\end{figure}
\begin{figure}[h!]
    \centering
    \includegraphics[width=\linewidth]{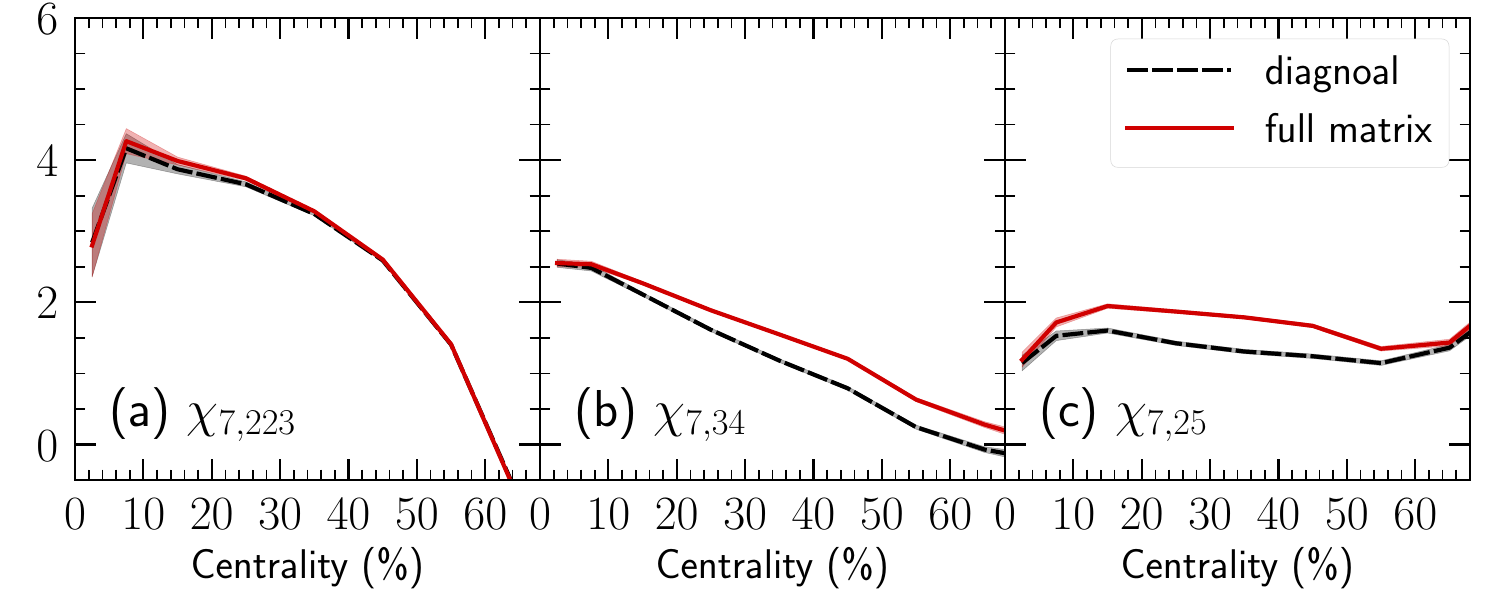}
    \caption{The nonlinear mode coefficients for $v_7$ by solving the full matrix equation vs. only including the diagonal elements.}
    \label{fig:chi7}
\end{figure}
\begin{figure}[h!]
    \centering
    \includegraphics[width=\linewidth]{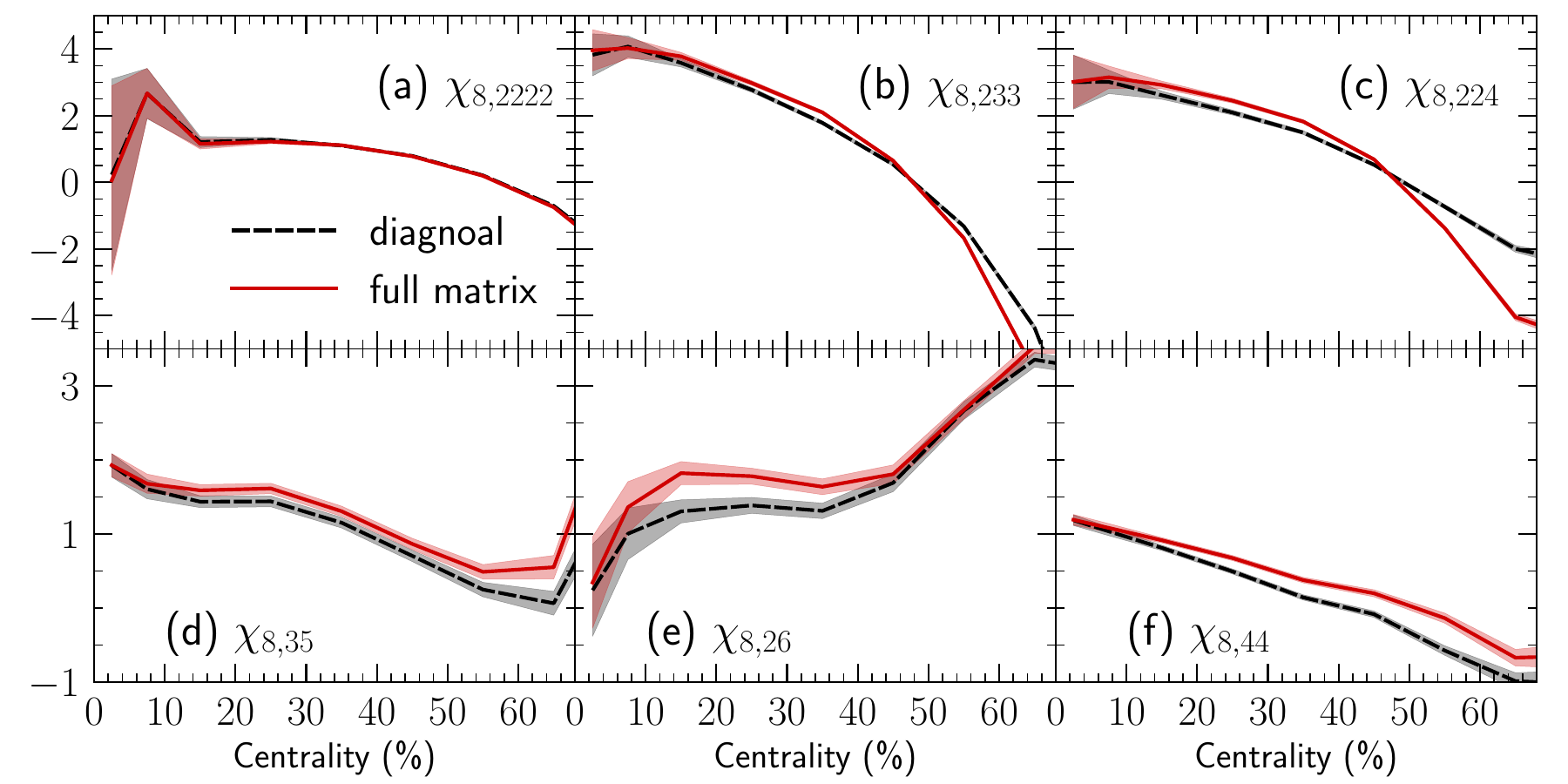}
    \caption{The nonlinear mode coefficients for $v_8$ by solving the full matrix equation vs. only including the diagonal elements.}
    \label{fig:chi8}
\end{figure}
\begin{figure}[h!]
    \centering
    \includegraphics[width=\linewidth]{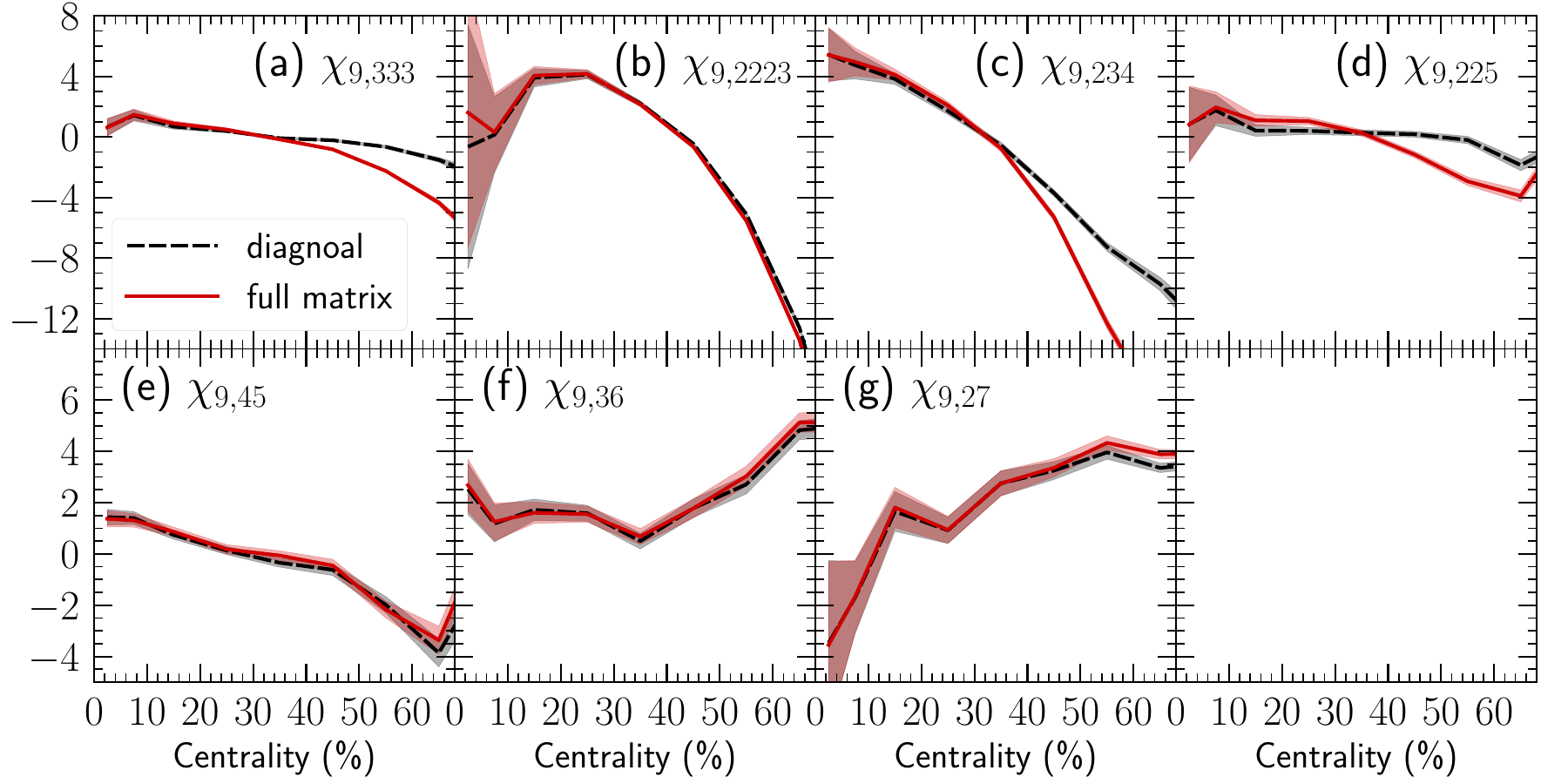}
    \caption{The nonlinear mode coefficients for $v_9$ by solving the full matrix equation vs. only including the diagonal elements.}
    \label{fig:chi9}
\end{figure}

Figures~\ref{fig:chi6} and \ref{fig:chi7} study the impact of such approximation on the nonlinear mode coefficients of $V_6$ and $V_7$. We find that the nonlinear mode coefficients $\chi_{6,222}$ and $\chi_{7,223}$ are not affected. Meanwhile, the values of $\chi_{6,33}$ and $\chi_{6,24}$ are larger when the off-diagonal elements are included. Based on Eq.~\eqref{eq:chi6_MatrixEq}, our results indicate that the $\langle (V_3^2)^* V_2 V_{4L} \rangle$ correlation gives a sizable contribution to the mode decomposition. For the mode decomposition for $V_7$, the coefficients $\chi_{7,34}$ and $\chi_{7,25}$ exhibit a similar situation.

Figures~\ref{fig:chi8} and \ref{fig:chi9} present our model predictions for the nonlinear mode coefficients for $V_8$ and $V_9$. They can be compared with future experimental measurements from the high-precision data. Unlike most nonlinear mode coefficients for $V_4$ to $V_7$, which are all positive below 60\% centrality, some coefficients are negative for $V_8$ and $V_9$ in semi-peripheral collisions.
Here, we also check the impact of solving the full matrix equation vs. only the diagonal components. We found the impacts from the off-diagonal elements are generally small for $V_8$'s and $V_9$'s nonlinear mode coefficients. Nevertheless, there are some sizable effects on $\chi_{8,224}$, $\chi_{8,44}$, $\chi_{9,333}$, $\chi_{9,234}$, and $\chi_{9,225}$ in semi-peripheral Pb+Pb collisions. 

\begin{figure}[t!]
    \centering
    \includegraphics[width=\linewidth]{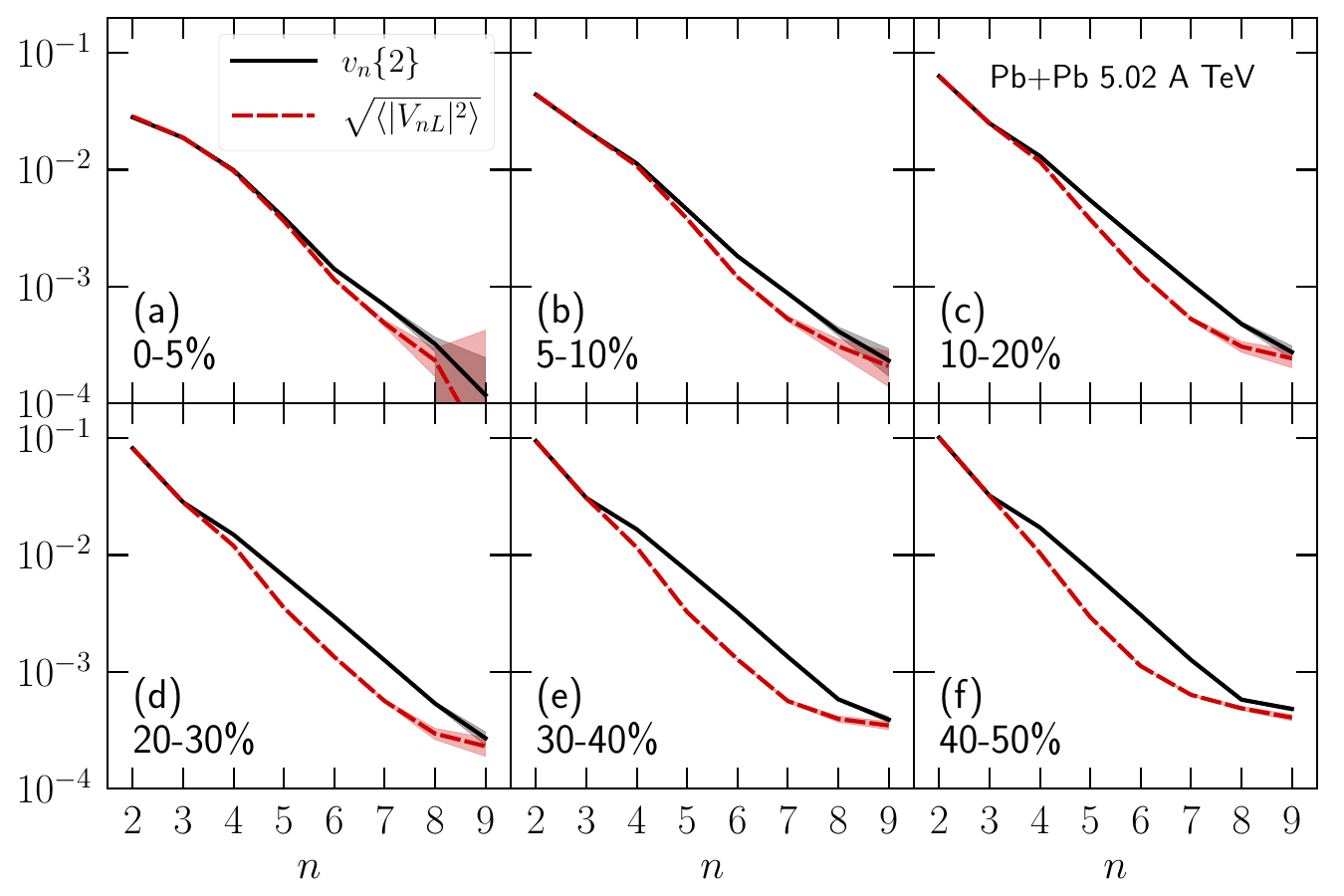}
    \caption{The RMS linear mode coefficients $\sqrt{\langle |V_{nL}|^2 \rangle}$ are compared with $v_n\{2\}$ in Pb+Pb collisions at $\snn = 5.02$~TeV.}
    \label{fig:vnL}
\end{figure}

\begin{figure}[t!]
    \centering
    \includegraphics[width=\linewidth]{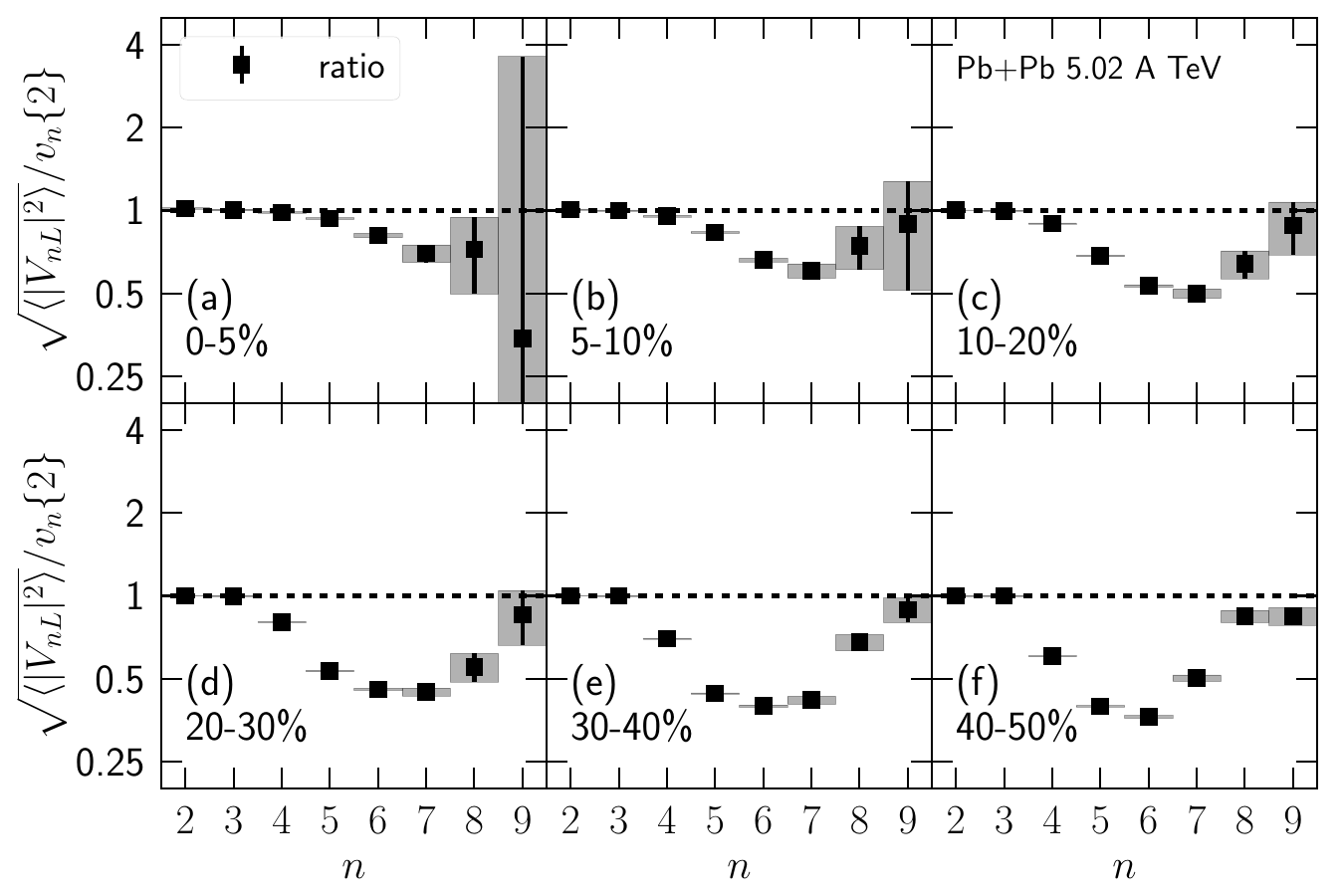}
    \caption{The ratio of the RMS linear mode coefficients $\sqrt{\langle |V_{nL}|^2 \rangle}$ over $v_n\{2\}$ in Pb+Pb collisions at $\snn = 5.02$~TeV.}
    \label{fig:vnLRatio}
\end{figure}

\begin{figure}[h!]
    \centering
    \includegraphics[width=\linewidth]{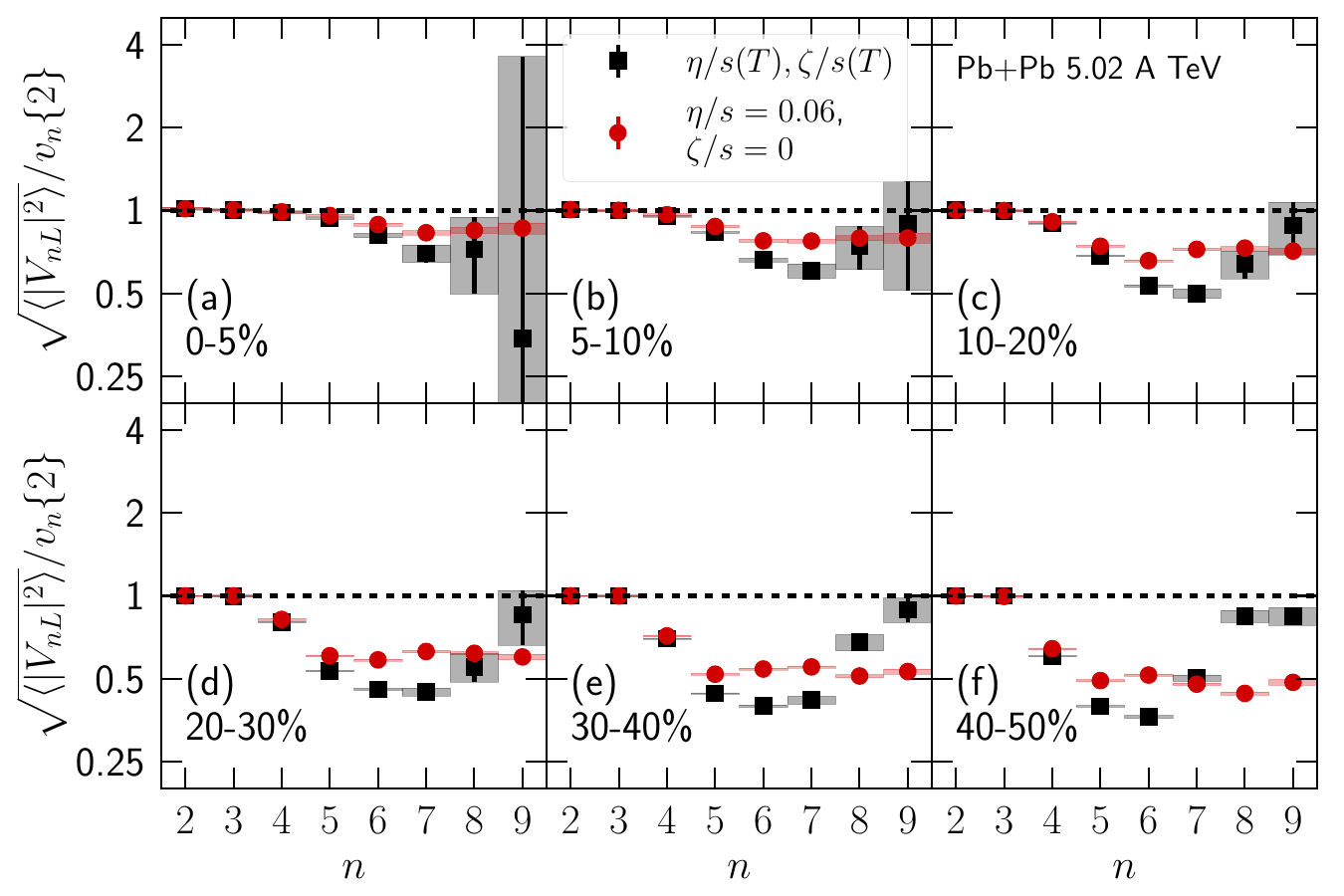}
    \caption{The ratio of the RMS linear mode coefficients $\sqrt{\langle |V_{nL}|^2 \rangle}$ over $v_n\{2\}$ with different QGP viscosities in Pb+Pb collisions at $\snn = 5.02$~TeV.}
    \label{fig:vnLRatioViscousEffects}
\end{figure}

After performing the mode decomposition, we obtain the event-by-event linear flow coefficients $V_{nL}$ for high-order anisotropic flow. Figures~\ref{fig:vnL} and \ref{fig:vnLRatio} compare their RMS values $\sqrt{\langle |V_{nL}|^2 \rangle}$ with the $v_n\{2\}$ in several centrality bins in Pb+Pb collisions at $\snn = 5.02$~TeV. In central collisions, we find that the linear mode coefficients play a dominant role in $ v_n\{2\} $ for all orders of $n$. In semi-peripheral collisions (20-50\%), the lowest-order elliptic flow coefficients are large because they receive contributions from the global collision geometry. Its related nonlinear terms become important for high-order anisotropic flow coefficients. Interestingly, Figure~\ref{fig:vnLRatio} shows the ratio $\sqrt{\langle |V_{nL}|^2 \rangle}/v_n\{2\}$ is non-monotonic as a function of $n$. The maximum mixture between the linear and nonlinear flow modes occurs at $n = 6$ in our model, at which the linear flow mode $\sqrt{\langle |V_{nL}|^2 \rangle}$ is about half of the $v_n\{2\}$. The nonlinear terms in the higher-order anisotropic flow coefficients $(n \geq 7)$ become subdominant in $v_n\{2\}$ because of the viscous damping on the nonlinear modes. Figure~\ref{fig:vnLRatioViscousEffects} shows that the ratios of $\sqrt{\langle |V_{nL}|^2 \rangle}/v_n\{2\}$ remain around 0.5 for all $n \geq 5$ when the QGP fluid is less viscous ($\eta/s = 0.06, \zeta/s = 0$) in the simulations. It will be interesting to verify such behavior in future experimental measurements.

\section{Conclusion}
\label{Sec:conclusion}

In this work, we studied the anisotropic flow power spectrum with the IP-Glasma + MUSIC + UrQMD framework. The high-statistics simulations enable us to push the state-of-the-art event-by-event hybrid hydrodynamic + transport framework to compute anisotropic flow coefficients up to $v_{12}\{2\}$ in high-energy nuclear collisions. The anisotropic flow power spectrum exhibits a strong sensitivity to the viscosity employed in the simulations. With the model calibration limited to elliptic flow measurements, we find that the initial-state fluctuation power spectrum predicted by the IP-Glasma model yields remarkable agreement with ALICE measurements for $v_2\{2\}$ up to $v_7\{2\}$, spanning central to 50\% centrality. Our results demonstrate that the IP-Glasma initial conditions capture fluctuations over a wide range of scales. In the meantime, the model underestimates the measured values of $v_8\{2\}$ and $v_9\{2\}$ in semi-peripheral collisions. 

We also performed systematic nonlinear mode decomposition for high-order anisotropic flow coefficients up to the harmonic order $n = 9$. Based on our model calculations, solving the full matrix equation is crucial for obtaining the correct values of some nonlinear mode coefficients for high-order anisotropic flow coefficients with harmonic order $n \geq 6$. 
It would be great to verify this difference in future experimental measurements.
We provide model predictions for the nonlinear mode coefficients for $V_8$ and $V_9$. We further compare the RMS of the linear flow components $\sqrt{\langle |V_{nL}|^2 \rangle}$ with $v_n\{2\}$. In semi-peripheral collisions, we find an interesting non-monotonic dependence of the ratios to $v_n\{2\}$ as a function of the harmonic order $n$ with the calibrated QGP viscosity in the simulations. The maximum mixture between the linear and nonlinear modes happens at $n = 6$. The nonlinear modes for $v_n$ with $n \ge 7$ are strongly damped by the viscosity compared to their linear modes. This finding is verified through simulations with low QGP viscosity, which show a significant nonlinear mode mixture for all high-order anisotropic flow coefficients. It will be interesting to search for these observations in the upcoming high-statistics experimental measurements. 

The data that support the findings of this article are openly available~\cite{shen_2025_15602042}, embargo periods may apply.

\section*{Acknowledgments}
We thank Hendrik Roch and Bj\"orn Schenke for fruitful discussion.
This work is supported by the U.S. Department of Energy, Office of Science, Office of Nuclear Physics, under DOE Award No. DE-SC0021969.
C.S. acknowledges a DOE Office of Science Early Career Award. 
This research was done using computational resources provided by the Open Science Grid (OSG)~\cite{Pordes:2007zzb, Sfiligoi:2009cct, OSPool, OSDF}, which is supported by the National Science Foundation awards \#2030508 and \#2323298.

\appendix
\section{Numerical convergence of $v_n\{2\}$}
\label{sec:numericalConvergence}

To test the numerical convergence of our model results for high-order anisotropic flow coefficients, we run simulations at three different grid spacings and rotate every collision event with random angles before hydrodynamic simulations.

\begin{figure}[h!]
    \centering
    \includegraphics[width=\linewidth]{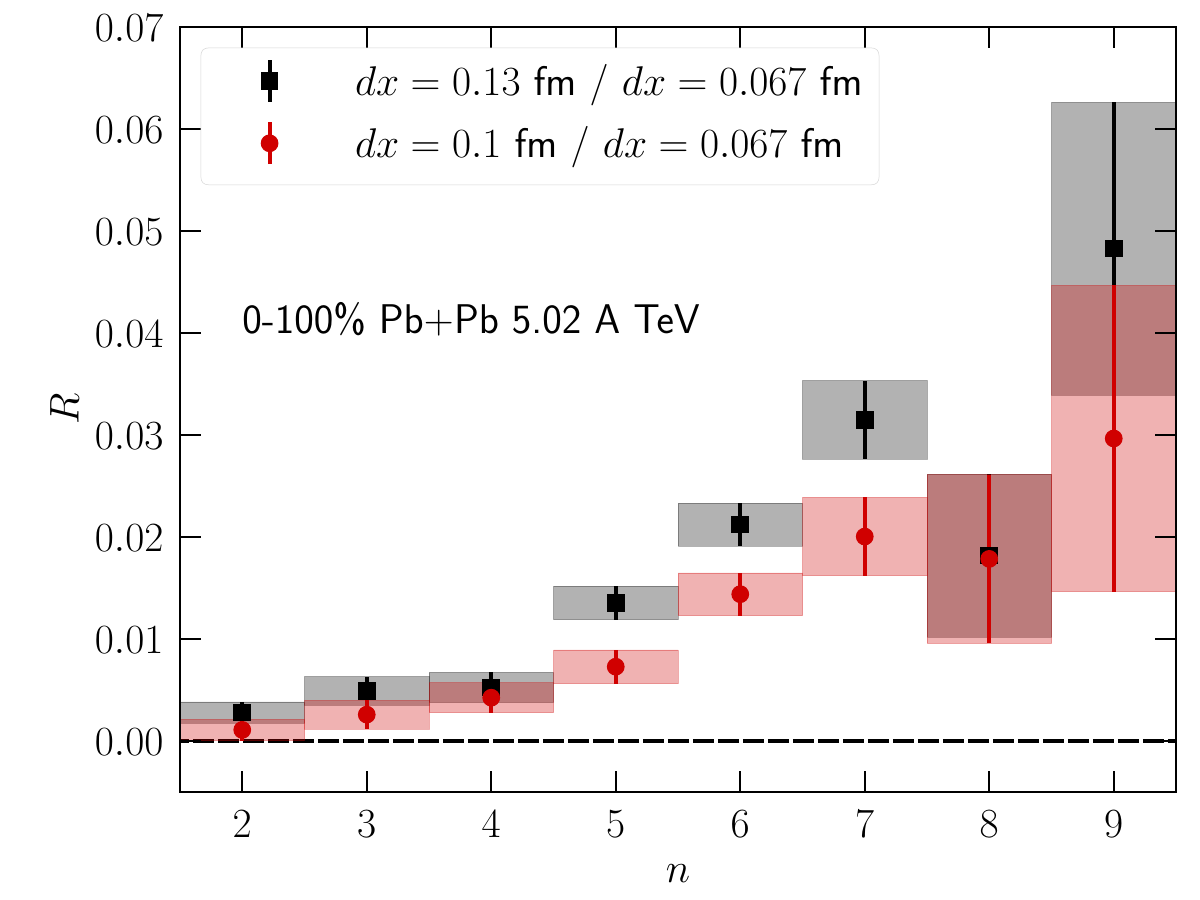}
    \caption{The relative difference in $v_n\{2\}$ from two sets of numerical simulations with different grid spacings $dx$ in the transverse plane.}
    \label{fig:GridSpacing}
\end{figure}

Figure~\ref{fig:GridSpacing} shows the relative difference between $v_n\{2\}$ computed with different grid spacings in hydrodynamic simulations. The relative difference between two observables $A$ and $B$ is defined as
\begin{align}
    R \equiv \frac{|A - B|}{A + B}. \label{eq:RelativeDiff}
\end{align}
Here, we test the numerical convergence of $v_n\{2\}$ using minimum-bias collision events to maximize the available statistics. With such high-statistics simulations, we observe that the relative difference in $v_n\{2\}$ approaches zero systematically with decreasing grid spacing. The relative errors grow roughly as $n^2$ for $v_n$ as a function of $n$. With a grid spacing of $dx = 0.067$ fm, the relative variation is smaller than 1\% for $v_n\{2\}$ with $n \le 5$ and 5\% for $v_9\{2\}$.

\begin{figure}[h!]
    \centering
    \includegraphics[width=\linewidth]{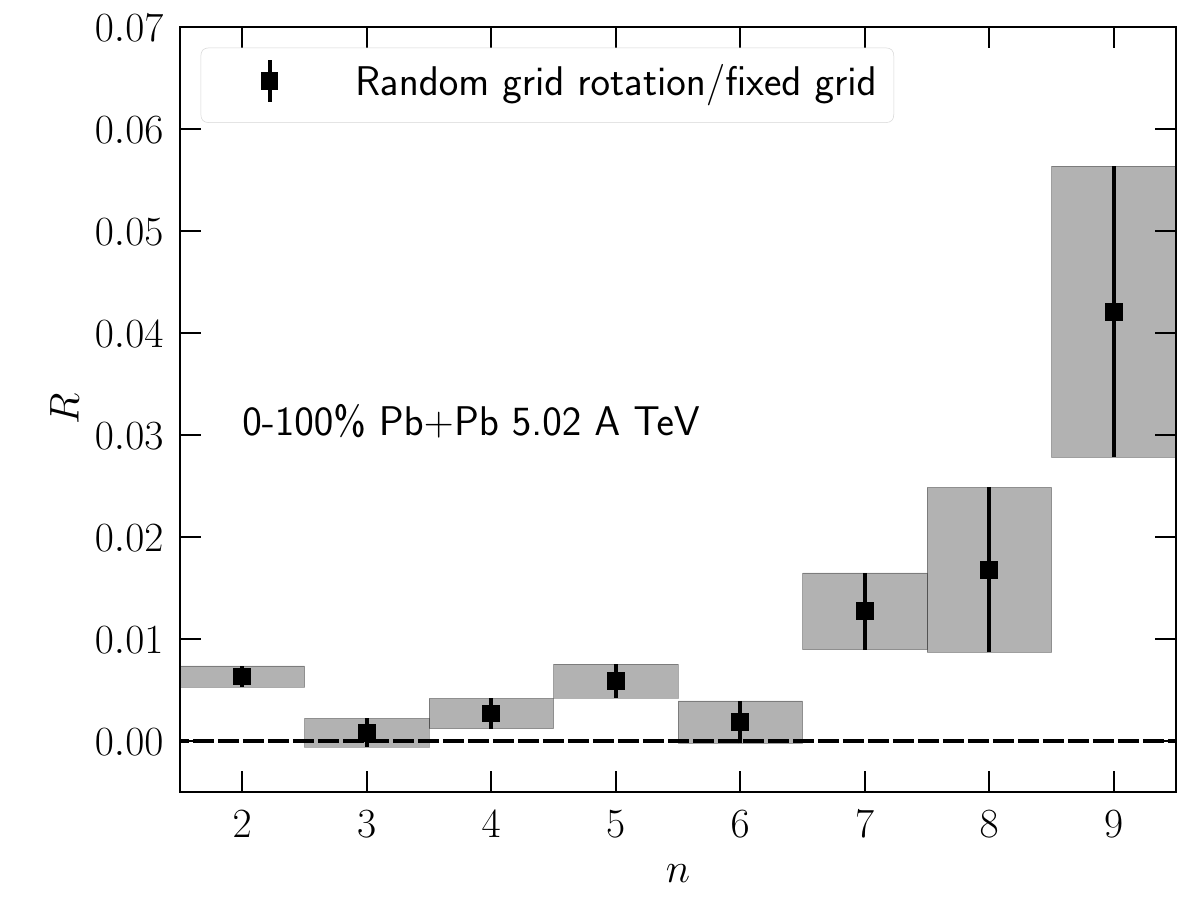}
    \caption{The relative difference in $v_n\{2\}$ from two sets of numerical simulations with and without random grid rotations on the initial conditions for hydrodynamics. The grid spacing is $dx = 0.067$ fm in the transverse plane.}
    \label{fig:GridRotation}
\end{figure}

Because hydrodynamic simulations utilize rectangular-shaped grids in the transverse plane, we perform convergence tests by rotating every collision event with random angles, ensuring that the rectangular grid is not aligned with any specific axis. Figure~\ref{fig:GridRotation} shows that the relative variation is below 1\% for $v_n\{2\}$ with $n \le 6$. The relative difference increases to $\approx 5\%$ for $v_9\{2\}$.

Based on these numerical tests, we can ensure that the relative numerical uncertainties in $v_n\{2\}$ are below 10\%.

\section{Decoupling of the directed flow $V_1$}
\label{sec:V1}

The flow decomposition framework in Sec.~\ref{Sec:analysis} starts with the elliptic flow $V_2$ as the lowest order flow coefficient. However, the lowest order anisotropic flow coefficient in the standard Fourier expansion is the directed flow $V_1$. Although the measured directed flow receives non-trivial contributions from global transverse momentum conservation on an event-by-event basis, the $V_1$ computed in the hybrid framework does not impose event-by-event correlations from these conservation laws. In this appendix, we will demonstrate that it is a good approximation to neglect nonlinear terms related to $V_1$ in the flow decomposition analysis. 

Considering directed flow in the nonlinear decomposition, we can start the mode analysis for elliptic and triangular flow coefficients as follows,
\begin{align}
    V_2 &= V_{2L} + \chi_{2,11}V_1^2 \label{eq:v2} \\
    V_3 &= V_{3L} + \chi_{3,12}V_1V_2. \label{eq:v3}
\end{align}
We can carry out nonlinear mode decomposition and obtain the linear flow coefficients $V_{2L}$ and $V_{3L}$. 

\begin{figure}[h!]
    \centering
    \includegraphics[width=\linewidth]{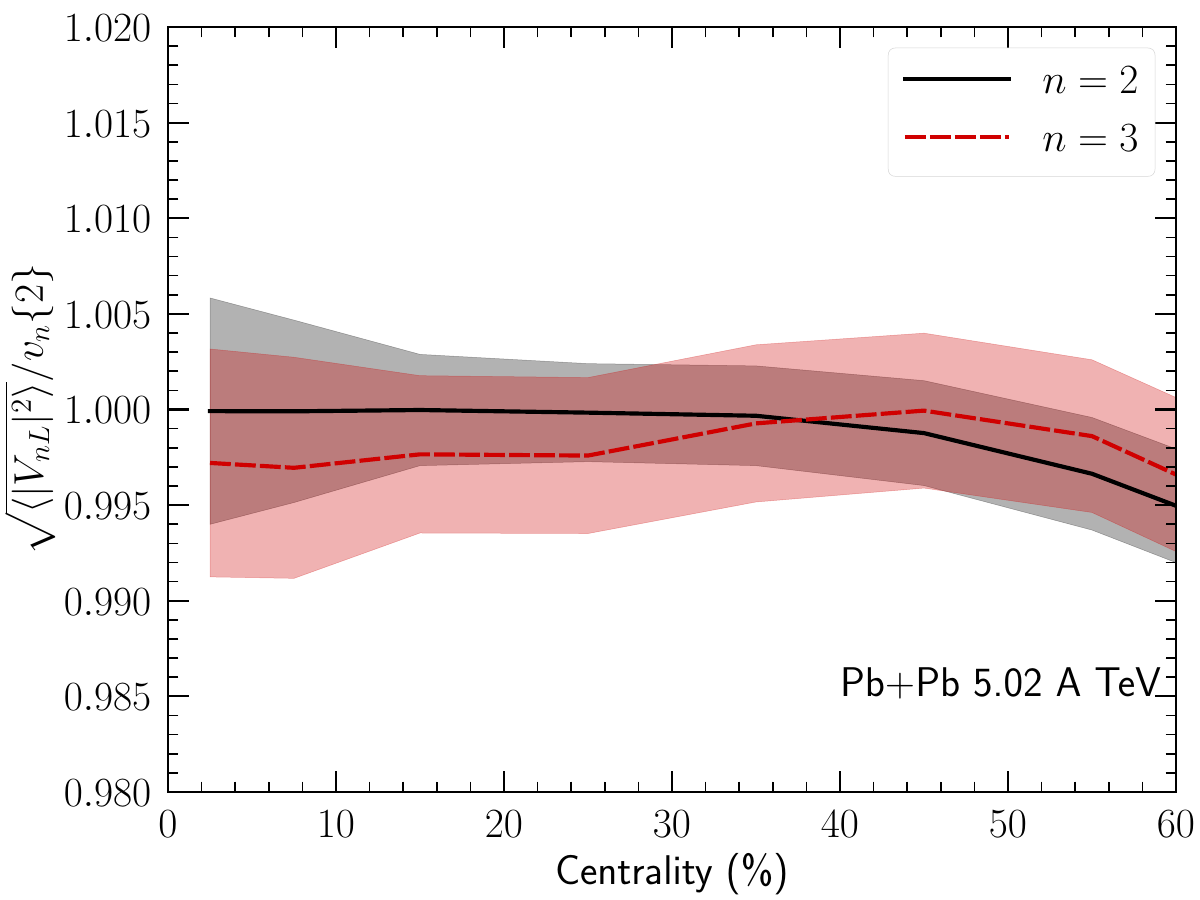}
    \caption{The ratio of the RMS $\sqrt{\langle |V_{nL}|^2 \rangle}$ and $v_n\{2\}$ for $n = 2, 3$ including the nonlinear terms from the directed flow $V_1$.}
    \label{fig:v2v3_v1}
\end{figure}

Figure~\ref{fig:v2v3_v1} shows the ratios of the RMS $\sqrt{\langle |V_{nL}|^2 \rangle}$ and $v_n\{2\}$ for elliptic and triangular flow. The ratios are extremely close to unity, indicating the linear flow mode dominates the flow, which is in line with our expectation that elliptic and triangular flow show good linear response to the initial-state eccentricities $\mathcal{E}_2$ and $\mathcal{E}_3$, respectively. 

To further examine the directed flow's impacts on $V_4$ and $V_5$, their decomposition equations are,
\begin{align}
    V_4 &= V_{4L} + \chi_{4,22} V_{2L}^2 + \chi_{4,13} V_1 V_{3L}  \label{eq:v4nonlinear} \\
    V_5 &= V_{5L} +\chi_{5,23} V_{2L} V_{3L} + \chi_{5,14} V_1 V_{4L} \label{eq:v5nonlinear}
\end{align}
Based on the results in Fig.~\ref{fig:v2v3_v1}, we can replace $V_{2L}$ and  $V_{3L}$ by $V_2$ and $V_3$, respectively. Similar to Eq.~\eqref{eq:chi6_MatrixEq}, we can solve the matrix equations for the nonlinear mode coefficients for $V_4$ and $V_5$ when the $V_1$ terms are present.

\begin{figure}[h!]
    \centering
    \includegraphics[width=\linewidth]{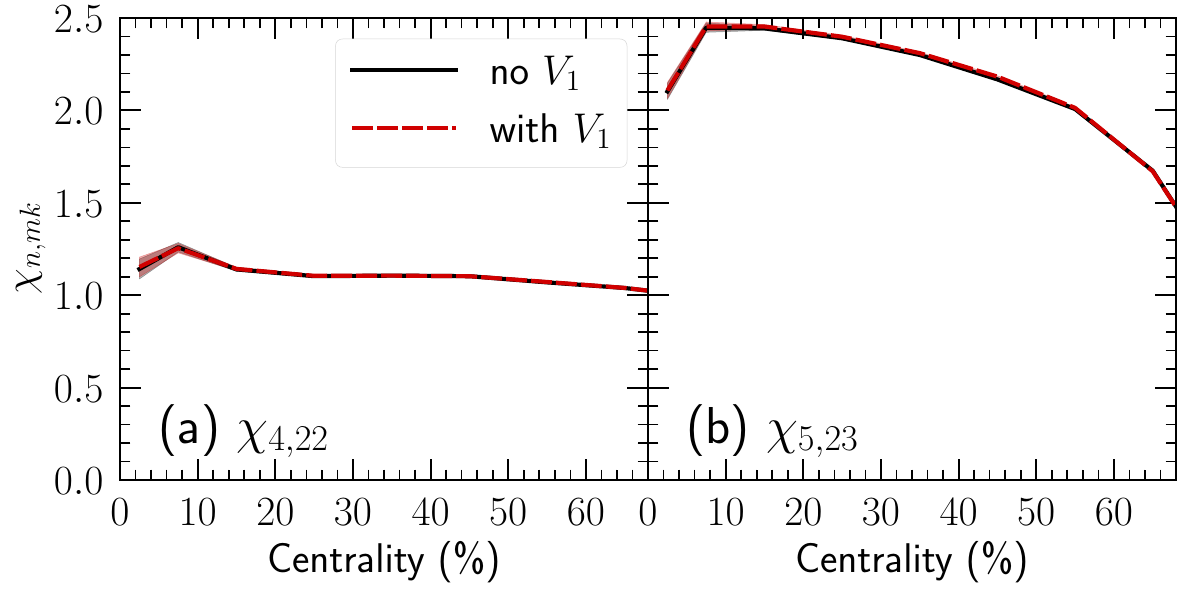}
    \caption{The nonlinear response coefficients for $v_4$ and $v_5$ with and without the nonlinear coupling terms involving the directed flow $V_1$.}
    \label{fig:chi45_v1}
\end{figure}

Figure~\ref{fig:chi45_v1} shows the extracted nonlinear mode coefficients $\chi_{4,22}$ and $\chi_{5,23}$ with and without including the $V_1$ terms in the mode decomposition. We find negligible effects from the off-diagonal components in this case. We also checked that the extracted linear flow coefficients $\sqrt{\langle |V_{nL}|^2 \rangle} (n = 4, 5)$ are unaffected with and without the $V_1$ terms.

In summary, we have numerically verified that the directed flow $V_1$ is decoupled from the high-order anisotropic flow coefficients in the nonlinear mode decomposition. It is a good approximation to ignore all the nonlinear flow terms related to $V_1$ in the mode expansion for anisotropic flow.

\bibliography{References, non-inspires}

\end{document}